# THE ASYMMETRIC DRIFT OF THE THICK DISK POPULATION

Devendra Ojha, Annie C. Robin and Olivier Bienaymé
*Observatoire de Besançon, BP 1615, F-25010 Besançon Cedex, France*

## ABSTRACT

We investigate the kinematics of the intermediate population using the photometric and astrometric sample surveys towards two opposite directions (galactic centre and anticentre) at intermediate latitude. A multivariate discriminant analysis (MDA) is used to distinguish the 'Thick Disk' from other populations with the help of the Besançon model of population synthesis. The data constrain the asymmetric drift of the intermediate population, which is found to be 40±10 km sec$^{-1}$ and do not depend on the galactic radius.

## 1  INTRODUCTION

Because the kinematics and metallicity of the thick disk is supposed to differ from the thin disk and halo, it may be possible to pick out thick disk stars on the basis of kinematical and/or metallicity differences between the three populations.

We therefore try here to find new constraints on thick disk population using samples at intermediate latitude which include photometry and proper motions. We used multivariate discriminant analysis (MDA) to qualify the thick disk using observations in a 5 dimensional space (V, B-V, U-B, $\mu_l$, $\mu_b$) and a model of population synthesis. The distribution of stars along the discriminant axis (combination of the observed axis) shows the clearest separation between the thin disk, thick disk and halo.

## 2  SUBSAMPLES

We have done the analysis by using two new data sets (Ojha et al. 1994a, 1994b) and the Besançon model of population synthesis (Robin & Crézé 1986; Bienaymé et al. 1987). The two data sets we used are the galactic centre field ($l = 3°$, $b = 47°$; area surveyed = 15.2 square degrees; hereafter GC) and galactic anticentre field ($l = 167°$, $b = 47°$; area surveyed = 8.6 square degrees; hereafter GAC). The overall accuracy in proper motions is 0."2 cen$^{-1}$. Figure 1 shows the comparison between 4 observed parameters (V, B-V, U-B, $\mu_l$) (solid line) with model predicted distributions (dotted line) towards GAC field. In the same figure, the thick disk population is overplotted by dotted-dashed lines. The 1 sigma error bar is $\pm\sqrt{N}$, where N is the number of stars in each bin. U-B and $\mu_l$ parameters are necessary to make a good discrimination between the three populations, because U-B is sensitive to the metallicity and $\mu_l$ is parallel to the V velocity and discriminate the populations by their asymmetric drift.

Using model predictions, we selected a subsample (where the majority of the thick disk stars exist), which we determined to be B-V$\leq$0.8 and 14$\leq$V$\leq$15.5 (see also figure 1). The observed counts per square degree of blue stars with above color and magnitude limits give 1893 stars in GC field, and 705 stars in GAC field.

In figures 2 & 3, we show the comparison between observed and model predicted $\mu_l$ distribution in the two fields for four different values of the circular velocity of thick disk. The effect of the choice of the asymmetric drift is well visible in these figures. It is quite clear from the figures that the $\mu_l$ parameter is very sensitive to the circular velocity of thick disk.

# 3 PROCEDURE

To avoid too large Poisson noise in the Monto Carlo simulations, we computed at least 10 simulations of 100 square degrees for each of the models tested in our analysis. The characteristics of each tested model are shown in Table 1.

We used the model simulations to find the best discriminant axes where to project the data, in order to separate the thick disk population from the disk and halo. This was done using a MDA (Multivariate Discriminant Analysis) under the MIDAS environment. The first discriminant axis for the circular velocity of 180 km sec$^{-1}$ of thick disk in the direction of galactic centre is given by :

$$x = 0.024(B - V) + 0.139(U - B) - 0.079V - 0.310\mu_l - 0.069\mu_b$$

It should be noted that the discriminant axis varies when we change the circular velocity in model simulations. The resulting discriminant axis is dominated by the proper motion along the rotation ($\mu_l$) and by the U-B

Table 1: **The characteristics of the Besançon model used for the present analysis. The Rotational velocity of the thick disk at the solar position ($V_{rot}$) was changed to produce the model simulated catalogues. The asymmetric drift for the thin disk was computed from velocity dispersion and density gradient**

|            | Age (in yrs) | $\sigma_U$ km s$^{-1}$ | $\sigma_V$ km s$^{-1}$ | $\sigma_W$ km s$^{-1}$ | Vrot km s$^{-1}$ | [Fe/H] |
|------------|--------------|------------------------|------------------------|------------------------|------------------|--------|
| Disk       | 0-.15E9      | 16.7                   | 10.8                   | 6.0                    | 218              | 0.01   |
|            | .15-1E9      | 19.8                   | 12.8                   | 10.0                   | 216              | 0.03   |
|            | 1-2E9        | 27.2                   | 17.6                   | 13.0                   | 213              | 0.03   |
|            | 2-3E9        | 30.2                   | 19.5                   | 18.5                   | 212              | 0.01   |
|            | 3-5E9        | 34.5                   | 22.2                   | 20.0                   | 207              | -0.07  |
|            | 5-7E9        | 34.5                   | 22.2                   | 20.0                   | 207              | -0.14  |
|            | 7-10E9       | 34.5                   | 22.2                   | 20.0                   | 207              | -0.37  |
| Thick disk | -            | 51.0                   | 38.0                   | 35.0                   | 180*             | -0.7   |
| Halo       | -            | 131.0                  | 106.0                  | 85.0                   | 0                | -1.7   |

color index due to metallicity differences between the thin disk, thick disk and halo.

## 4 STATISTICAL TESTS

**Kolmogorov-Smirnov test :** To quantitatively estimate the adequacy of the models with various circular velocities, we first applied Kolmogorov-Smirnov test to the cumulative distributions of model predicted stars and the observed one on the first discriminant axis. The results obtained are shown in Table 2. In both cases, we obtain the most probable value for the circular velocity of thick disk around 180 km sec$^{-1}$, that is a corresponding lag or asymmetric drift of 40±10 km sec$^{-1}$ at 2 sigmas with respect to LSR (Local Standard of Rest).

$\chi^2$ **test :** We applied another statistical test ($\chi^2$ test) to compare the distribution of the sample on the discriminant axis with a set of model predicted distribution, assuming different circular velocities of thick disk. $\chi^2$ test is more efficient in our present analysis compared to KS test. The $\chi^2$ distribution is given by a simple formula as follows:

$$\chi^2 = \sum_{i=1}^{n} \frac{(b_i - a_i)^2}{a_i}$$

Table 2: **Kolmogorov-Smirnov test (Probability of each model to come from the same distribution as the data). Models differ by their circular velocity of thick disk. The corresponding lag (asymmetric drift) is with respect to the LSR, assuming $V_{LSR} = 220$ km sec$^{-1}$ (IAU 1985) and $V_{\odot}=11$ km sec$^{-1}$ (Delhaye 1965)**

| $V_{cir}$ (km sec$^{-1}$) | Lag (km sec$^{-1}$) | Probability GC | GAC |
|---|---|---|---|
| 140 | 80 | 0.12 | 0.01 |
| 160 | 60 | 0.29 | 0.01 |
| 173 | 47 | 0.40 | 0.05 |
| 180 | 40 | 0.77 | 0.30 |
| 190 | 30 | 0.06 | 0.05 |

Where n is the number of bins and *a* & *b* are the number of counts in each bin in the model and observed data set, respectively. We use an approximation (in number of sigmas) to the distribution of $\chi^2$ given by a function (Kendall & Stuart 1969):

$$\sigma_{\chi^2} = \{(\frac{\chi^2}{n-1})^{1/3} + \frac{2}{9(n-1)} - 1\}(\frac{9(n-1)}{2})^{1/2}$$

Table 3 and figure 4 show the values of the probability (in $\sigma_{\chi^2}$) of each model to come from the same distribution as the observed sample. The most probable value of circular velocity for the thick disk comes out to be 180 km sec$^{-1}$ (see figure 4). We notice that the model predictions are at 3 sigmas of the data. This is due to the fact that the statistics of the errors in the data is not a Poisson statistics, because of some systematic errors in the photometry.

The distribution over the discriminant axis of observed stars (full line) towards GAC ($l = 167°$, $b = 47°$) with best model predictions (with three populations) is shown in figure 5.

# 5 CONCLUSION

We have performed an attempt to analyse 5D data in order to discriminate the thick disk and to measure its asymmetric drift in two opposite directions.

Table 3: $\chi^2$ test for models with different circular velocities of thick disk. Lag or asymmetric drift is with respect to the LSR, assuming $V_{LSR}$=220 km sec$^{-1}$ (IAU 1985)

| $V_{cir}$ (km sec$^{-1}$) | Lag (km sec$^{-1}$) | $\chi^2$ (in sigmas) | |
|---|---|---|---|
| | | GC | GAC |
| 150 | 70 | 8.7 | 6.8 |
| 165 | 55 | 6.8 | 5.7 |
| 175 | 45 | 4.1 | 3.8 |
| 180 | 40 | 4.3 | 3.4 |
| 185 | 35 | 4.9 | 3.5 |
| 190 | 30 | 5.5 | 4.0 |
| 215 | 5 | 6.7 | 5.1 |

The present analysis estimates the circular velocity of the thick disk to be 180±10 km sec$^{-1}$. Assuming the Local Standard of Rest $V_{LSR} = 220$ km sec$^{-1}$, this gives an asymmetric drift of the thick disk of 40±10 km sec$^{-1}$ with respect to the LSR. We obtained a unique value in both directions, showing that no radial gradient seems to occur on a base of 3 kpc around the Sun. It should be noted that this analysis is not sensitive to the scale height and the velocity dispersion used for the thick disk.

The previous determinations of the asymmetric drift of thick disk are given as : Wyse & Gilmore (1986) - 100 km sec$^{-1}$, Norris (1987) - 20 km sec$^{-1}$, Sandage & Fouts (1987) - 50 km sec$^{-1}$, Ratnatunga & Freeman (1989) and Carney et al. (1989) - 30 km sec$^{-1}$, Spaenhauer (1989) - 80 km sec$^{-1}$, Morrison et al. (1990) - 35 km sec$^{-1}$, Soubiran (1993) - 52 km sec$^{-1}$ and Ojha et al (1994a, 1994b) - 50 km sec$^{-1}$. In our recent determinations (Soubiran 1993; Ojha et al. 1994a & 1994b), we do not found the dependence of the thick disk asymmetric drift with z- distance. Our approach allows to avoid three biases, first, by using a model of population synthesis, second, by adopting same selections in the model and data, and third, by identifying the thick disk from other populations on a physical basis, by its metallicity and kinematics. It is suggested that the differing characteristics of the thick disk found in the literature are due to different selection biases in the stellar samples but also to differing difficulties in separating the thick disk from the other stellar populations.

# Caption to the figures

**Figure 1.** Comparison between observed and model predicted histograms (V, B-V, U-B, $\mu_l$) towards GAC direction ($l = 167°$, $b = 47°$). Data (solid line), model predictions (dotted line) and model predicted thick disk (dotted-dashed line).

**Figure 2.** Comparison between observed and model predicted $\mu_l$ histograms towards GAC direction ($l = 167°$, $b = 47°$) for four values of the circular velocity of thick disk. Data (solid line), model predictions (dotted line) and model predicted thick disk (dotted-dashed line).

**Figure 3.** Comparison of observed and model predicted $\mu_l$ histograms towards GC direction ($l = 3°$, $b = 47°$) for four values of the circular velocity of thick disk. Data (solid line), model predictions (dotted line) and model predicted thick disk (dotted-dashed line).

**Figure 4.** $\chi^2$ test for models of different circular velocities of thick disk for the two fields. $\chi^2$ is given in number of sigmas.

**Figure 5.** Distribution over discriminant axis of observed stars (full line, with 1 sigma error bars) towards GAC field, best model predictions (dotted line) and model predicted stars according to their populations (dashed line : disk, dashed-dotted : thick disk and dahed-dotted-dotted : halo).